\documentclass[conference]{IEEEtran}
\IEEEoverridecommandlockouts
\usepackage{cite}
\usepackage[table]{xcolor}
\usepackage{multirow}
\usepackage{amsmath,amssymb,amsfonts}
\usepackage[caption=false]{subfig}
\usepackage{booktabs}
\usepackage{float}
\usepackage{algorithmic}
\usepackage{graphicx}
\usepackage{tabularx}
\usepackage{textcomp}
\usepackage{xcolor}
\usepackage{makecell}
\makeatletter
\usepackage{graphicx}
\DeclareRobustCommand*{\IEEEauthorrefmark}[1]{%
 \raisebox{0pt}[0pt][0pt]{\textsuperscript{\footnotesize #1}} }
\def\BibTeX{{\rm B\kern-.05em{\sc i\kern-.025em b}\kern-.08em
    T\kern-.1667em\lower.7ex\hbox{E}\kern-.125em}}
\providecommand{\keywords}[1]{\textbf{\textit{Keywords---}} #1}

\title{\LARGE \bf
Leveraging Open Threat Exchange (OTX) to Understand Spatio-Temporal Trends of Cyber Threats: Covid-19 Case Study
}
\author{
\IEEEauthorblockN{Othmane Cherqi
\IEEEauthorrefmark{1} \IEEEauthorrefmark{2}, Hicham Hammouchi \IEEEauthorrefmark{1}
\IEEEauthorrefmark{2}, Mounir Ghogho
\IEEEauthorrefmark{1}, Houda Benbrahim\IEEEauthorrefmark{2}}
\IEEEauthorblockA{
\IEEEauthorrefmark{1} Universit\'e Internationale de Rabat, 
Faculty of Informatics and Logistics, TICLab \\
\IEEEauthorrefmark{2} Mohammed V University, ENSIAS \\
Emails: \{othmane.cherqi, hicham.hammouchi, mounir.ghogho\}@uir.ac.ma, benbrahimh@hotmail.com} 
}

\begin{document}
\graphicspath{ {./images/} }
\maketitle
\thispagestyle{empty}
\pagestyle{empty}

\begin{abstract}
Understanding the properties exhibited by Spatial-temporal evolution of cyber attacks improve cyber threat intelligence. In addition, better understanding on threats patterns is a key feature for cyber threats prevention, detection, and management and for enhancing  defenses. In this work, we study different aspects of emerging threats in the wild shared by 160,000 global participants form all industries. First, we perform an exploratory data analysis of the collected cyber threats. We investigate the most targeted countries, most common malwares and the distribution of attacks frequency by localisation. Second, we extract attacks' spreading patterns at country level. We model these behaviors using transition graphs decorated with probabilities of switching from a country to another. Finally, we analyse the extent to which cyber threats have been affected by the COVID-19 outbreak and sanitary measures imposed by governments to prevent the virus from spreading.

\end{abstract}

\section{Introduction}

Cyber threats and vulnerabilities are constantly emerging and are becoming complex and diverse with the evolution of technology. These emerging threats can lead to devastating attacks causing loss of confidential data, money and reputation damage \cite{johns2020cyber}. In addition, due to the current covid-19 pandemic witnessed an explosion in the number of cyber attacks as most activities moved to operate online, affecting both individuals and organizations. This urges organizations to work towards a community driven approach to enable a strong exchange of information, knowledge and intelligence about these threats. In this sense, organizations are joining forces and sharing Cyber Threat Intelligence (CTI) through various platforms such as CINS score\footnote{https://cinsscore.com/}, Blocklist.de\footnote{http://www.blocklist.de/en/},
Abuse.ch\footnote{https://abuse.ch/}, and Alienvault OTX\footnote{https://otx.alienvault.com/}. 

As organizations increasingly adhere to this process, the number of events and incidents shared is increasing exponentially. Hence, automated mining and analysis of information on these platforms is of vital importance. Understanding the specifics of shared information and turning them into actionable insights would help organizations anticipate cyberattacks and mitigate their adverse effects \cite{wagner2019cyber}.  

 \textit{Cyber Attacks in the Era of Covid-19:} The fast spread of COVID-19 has led to the use of certain technologies in a rapid and unusual manner. E-learning and home-office, for example, have increased the number of users of certain platforms such as Zoom and Microsoft Team as well as other video conferencing softwares. This sudden surge of users has created an environment favorable to the spread of cyber attacks.

 In this work, we leverage information and intelligence covering different threats targeting various countries shared on AlienVault Open Threat Exchange (OTX). OTX is one of the largest platforms used to share cyber threats. By the time of our study, this platform contained 155K shared threats and 160K global users from 144 different countries. We explore different trends and patterns related to specific countries and malware families. We discuss campaign (mass) and targeted attacks, explore clusters of countries with the same characteristics, and analyze spatial-temporal trends of these threats and their spreading patterns across countries/organizations. In addition, analyze in depth the period of COVID-19 and investigate the most impacting threats. 
 
 This work attempts to answer several questions such as:
\begin{itemize}
    \item What are the most targeted countries and common used malware families ? 
    \item How these attacks spread between countries?
    \item What are the main differences between the the pre-covid and covid era cyberspace?
\end{itemize}

This type of analysis would help organizations better understand and anticipate threats. Integrating this information would help them understand the different techniques cybercriminals use against them based on the different profiles established. In addition, the more knowledge organizations have about the threats in the covid era, the better they can strengthen their defenses in the post-covid era.

The present paper is structured as follows: Next section provides a review of the related work to this study. Section \ref{Attacks Exploratory Analysis} presents an exploratory analysis of the cyber attacks. Section \ref{DataSet} is a presentation of the data set used. In section \ref{Clusterin}, we identify countries clusters. Finally, in section \ref{covid-19}, we analyse the extent to which cyber threats trend have been affected by the COVID-19 outbreak and sanitary measures imposed by governments. Section \ref{conlusion} concludes the paper.

\section{Related Work}\label{Related Work}

Cyber attacks targeting industrial, companies and governmental organizations have become widespread and several attacks have made headlines in the last decade\cite{guardian}. Recent cyber attacks have demonstrated the growing evolution of security incidents and the immense damage they cause: loss of intellectual property, productivity, money and reputation\cite{impact1}\cite{impact2}. Security professionals are increasingly willing to share information about the current threats and attacks they face. This information sharing appears to be a promising form of early warning\cite{collab}, which can be used to take preventive measures to protect information systems. It is not surprising that both researchers and practitioners have taken the initiative to create methods and build tools that would allow the automated exchange of security information and use it to increase the level of protection and threat anticipation. 

Cyber Threat Intelligence (CTI) has become a hot topic and researchers have worked on analyzing sharing platforms content. Abu et al\cite{cti1} identified several issues for threat data quality on sharing platform. The study includes a survey of common language and tools available in CTI. Sauerwein et al.\cite{cti2} proposed an exploratory study of software vendors of threat intelligence sharing platforms. The study identified a growing interest towards threat intelligence sharing and a lack on the consistence of shared threats. Gschwandtner et al.\cite{cti3} proposed a framework for integrating threat intelligence within existing information systems. They evaluated the framework and its reliability with information security practitioners by studying its ability to generate a vulnerability report, prioritizing critical vulnerabilities and related mitigation activities.

Few studies explored the spatio-temporel dependencies of cyber threats. Dowling et al.\cite{spatemp1} analyzed temporal and spatial  large-scale
cyber attacks. The paper demonstrate a favorable correlation between the recorded activity and the established probability. On the other hand Du et al.\cite{spatemp2} proposed a work flow for identifying attack sources belonging to the same coordinated attack groups. 

Our work generalizes the approach in \cite{spatemp1,spatemp2} by constructing graphs for each spatial source of attacks (IPs) over time, in order to uncover the general threat patterns followed by attackers, and to predict incidents rates at the country level. 

\textit{Cyber Security in the Age of COVID-19: }The COVID-19 crisis is associated with drastic, unprecedented changes in people's daily live, especially related to the use of multimedia (E-learning, home office...).This can lead to new crime opportunities\cite{eian2020cyber}. Online fraud and cyber-attacks have considerably increased during the COVID-19 pandemic. Cyber-threats rates have been particularly high during the periods were the lockdown policies were the strictest\cite{buil2021cybercrime}. Also during the pandemic, cyber criminals have used Advanced Persistent Threat (APT) targeted vulnerable user and systems taking advantage of the new domestic use of technology \cite{pranggono2021covid}\cite{lallie2021cyber}.  
 
\section{Data Set}\label{DataSet}
\subsection{Data Collection}\label{collection}
Open Threat Exchange (OTX) is a free to use crowd-sourced computer-security platform. It is the world largest platform of this kind with more than 160K participants in 144 countries. This data set contains incidents that targeted particular industries and organizations in different regions of the world. We developed our own web crawler and stored all the incidents as raw HTML pages. We were able to extract over than 13000 unique security events occurred between 2015 and Sep. 2020.

\subsection{Data Preprocessing}

In this section we will briefly describe how  we  reduced  scraped HTML files into a structured  form containing  the  relevant information about attacks. We used the semi structured form of HTML and the tags specific to each field to be able to find more easily the information we needed on each web page. The main information we have extracted about the incidents are: incident description, incident title, targeted country, date of the incident and the adversary responsible of the attack.
As our goal is to predict security incidents' rates over governments, we need a geographical view outlining the evolution and transition of attacks from one country/state to another, which we find in this dataset.
The parsed pages containing shared threats as entered by the stakeholders need to be processed since countries names are usually misspelled. In addition, since our approach is geographically driven, we need to unify countries name spelling. To this end, we the Levenshtein distance \cite{levenshtein} to compare two strings as given by Equation \ref{eq:algo_lev}:
\\
\begin{equation}
\label{eq:algo_lev}
\resizebox{0.9\hsize}{!}{$%
 lev_{a,b}(i,j) = \left\{\begin{matrix}
max(i,j) & if min(i,j)=0\\ 
min\left\{\begin{matrix}
 lev_{a,b}(i-1,j)+ 1 & \\ 
 lev_{a,b}(i,j-1)+ 1 & \\ 
 lev_{a,b}(i-1,j-1)+ 1_{(a_{i}\neq b_{j})} & \\
\end{matrix}\right. & otherwise. 
\end{matrix}\right.
$%
}%
\end{equation}
\\

This distance measures the difference between two strings by examining the number of changes (i.e. insertions, deletions or substitutions) made to a character to change one word into another. This approach is widely used in text mining and Natural Language Processing fields\cite{levenshtein_nlp}. We used a dictionary containing all the country names written correctly and use the algorithm above to obtain a data set with homogeneous and correct country names.

\section{Attacks Exploratory Analysis}
\label{Attacks Exploratory Analysis}

In this section, we analyze the collected incidents data in order to gain insight into the underlying patterns such as which countries are most targeted, the frequency of attacks, and how these attacks spread from country to country. 
\subsection{Most Targeted Countries}
\label{mst_targetd}
\begin{figure}[ht]
\centering
\includegraphics[width=\linewidth]{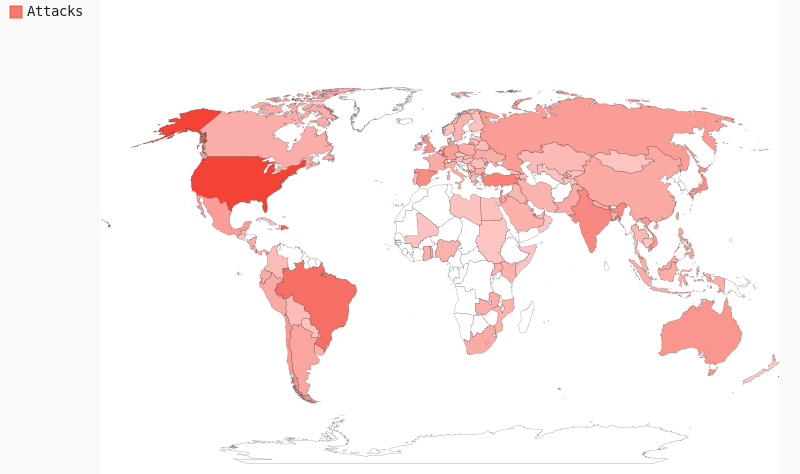}
\caption{Heat map of the reported cyber incidents over countries (2016-2020)}
\label{fig:map_attacks}
\end{figure}

\begin{figure}[ht]
\includegraphics[scale=0.6]{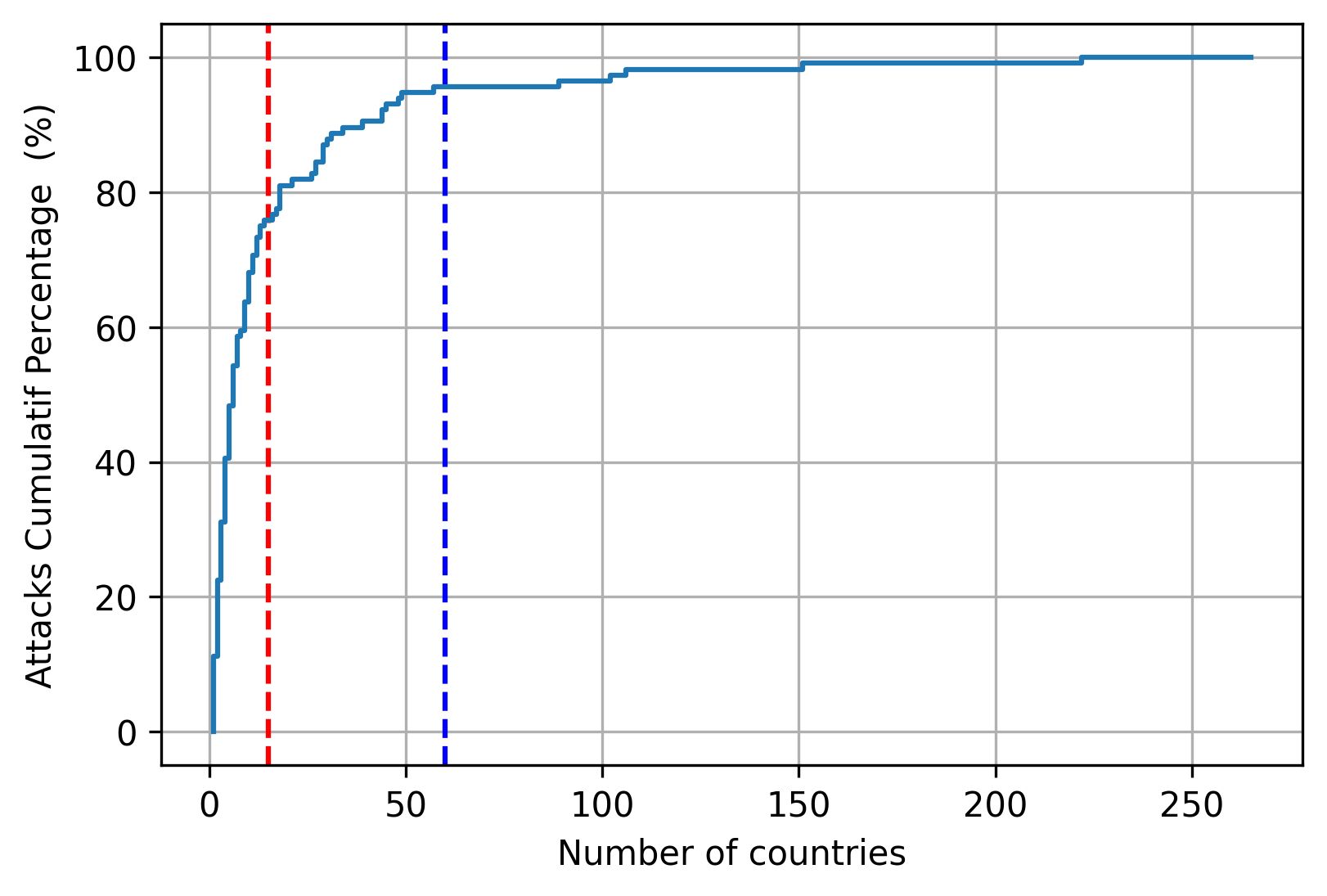}
\caption{Attacks cumulative percentage by countries)}
\label{fig:cumsum}
\end{figure}

We begin by aggregating the attacks reported for each country. That is, we aggregate the threats posted by country name and we count the number of incidents for each one. Fig. \ref{fig:map_attacks} shows a heatmap of the most attacked countries (dark color) to the less attacked countries. Also, we observe that the most targeted countries are, as excepted, USA, China and Russia alongside. We can conclude that attackers tend to target most powerful countries known as super power: USA, China, Russia. European Union (E.U) countries combined were hit by 31\% of all the attacks. United States is subject to nearly 22.2\% of the attacks. Furthermore, among 140 countries, only 17 attract 80\% of the threats and 45 countries attract 99\% of it as shown by the cumulative distribution function in  Figure \ref{fig:cumsum}.

\subsection{Attacks Evolution}

\begin{figure}[ht]
\includegraphics[scale=0.6]{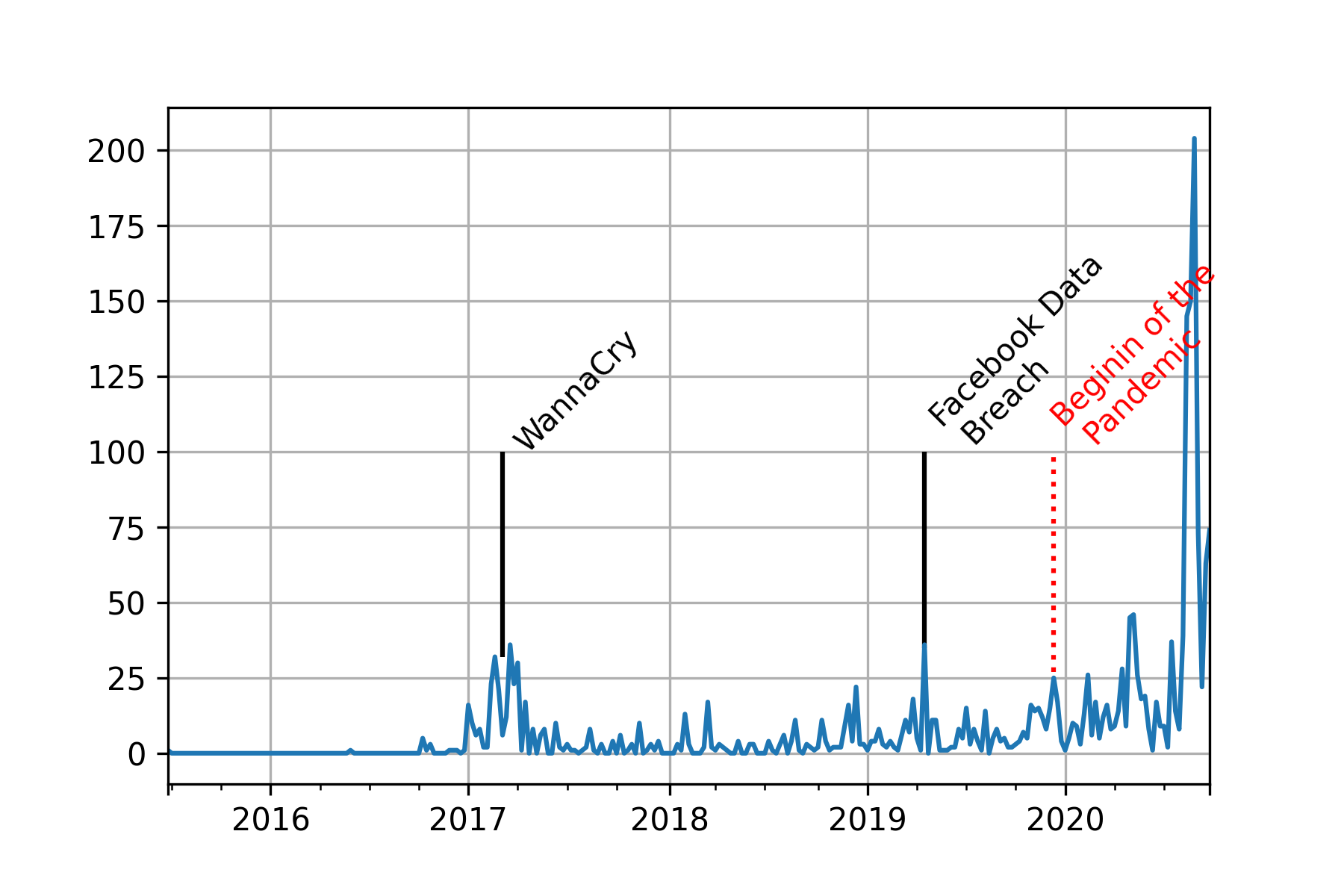}
\caption{Threats evolution's time series (2015-2020)}
\label{fig:timeseries}
\end{figure}

In order to study the frequency of attacks between 2016 and 2020, we construct a time series of number of attacks. As shown in fig. \ref{fig:timeseries}, we observe that the time series of the attacks is quite stationary except for the two peaks in March and August 2017, and slight increase in January and April 2019. The two spikes of 2017 could be explained by two major attacks, namely,  WannaCry in May 2017 and the Apache Struts vulnerability in March 2017. As for the increase in 2019, this coincides with the Facebook user data leaks (wich came almost exactly one year after the Cambridge Analytica scandal). Also, the big increase in 2020 could be highly related to the covid-19 pandemic and its governmental measures.

In the same vein, looking at the most used malware families illustrated in the table \ref{importances}, we observe that 2019 was characterized by ransomware attacks that are primarily motivated by financial gain. As for 2020 and with the pandemic, the attacks began to shift from financial gain to information seeking by using spyware to obtain sensitive information.

\begin{table}[ht]
     \centering
     \caption{Top 5 most common malware families in the last 3 years}
     \label{importances}
     \begin{tabular}{|c|c|c|}
     \hline 
       Year & Top malwares & Type \\ \hline \hline
       \multirow{5}{*}{\makecell{2019}} &  Emotet & Banking trojan \\
       & trojan:win32/nukesped & Ransomware \\ 
       & Dridex & Banking trojan\\
       & ransom:win32/wannacrypt & Ransomware \\
       & trojan:win64/turla & Trojan Spy (espionage) \\
       \hline
       \multirow{5}{*}{\makecell{2020}} &  Emotet & Banking trojan\\
       & Phishing  & Phishing \\
       & backdoor:php/webshell & Backdoor\\
       & trojan:win32/qakbot &  Backdoor trojan\\ 
       & w97m/downldr.ie.gen!eldorado & MS Word macro trojan\\
       \hline
       \multirow{5}{*}{\makecell{2021}} &  Cobalt Strike & Penetration tool \\
       & fp539598-vbs/loveletter.bt  & Mass-mailing worm \\
       & revil & Ransomware\\
       & win32:dangeroussig & Trojan \\ 
       & win.packed.agentino-9863792-0 & Trojan\\
       \hline
     \end{tabular}
 \end{table}

\subsection{Attacks Spreading Pattern }

In this section, we model the spreading of cyber attacks using transition probability matrix to generate transition graphs. We assess the relationship between countries by determining the probability of transition of the same attack from a country to another, then, we identify ways attacks spread geographically.
\subsubsection{Most Targeted Tuples}
\begin{figure}[ht]
\includegraphics[scale=0.6]{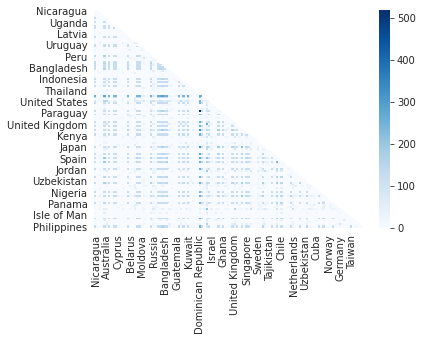}
\caption{Most frequent attacked pairs of countries} 
\label{fig:pairs_attacks}
\end{figure}

Here we try to identify the country pairs that often appear in the same attack campaigns. Thus the Figure \ref{fig:pairs_attacks} shows that it is  generally neighbouring countries that are often affected by the same attacks. However, we also see that the United States is also linked to several countries. This is explained by the fact that it is by far the most attacked country as seen in Section \ref{mst_targetd}.
\subsubsection{Attacks' Transition Graph}
\begin{figure}[!h]
  \centering
  \subfloat[Targeted attack pattern on Cluster 0.]{\includegraphics[width=0.47\textwidth]{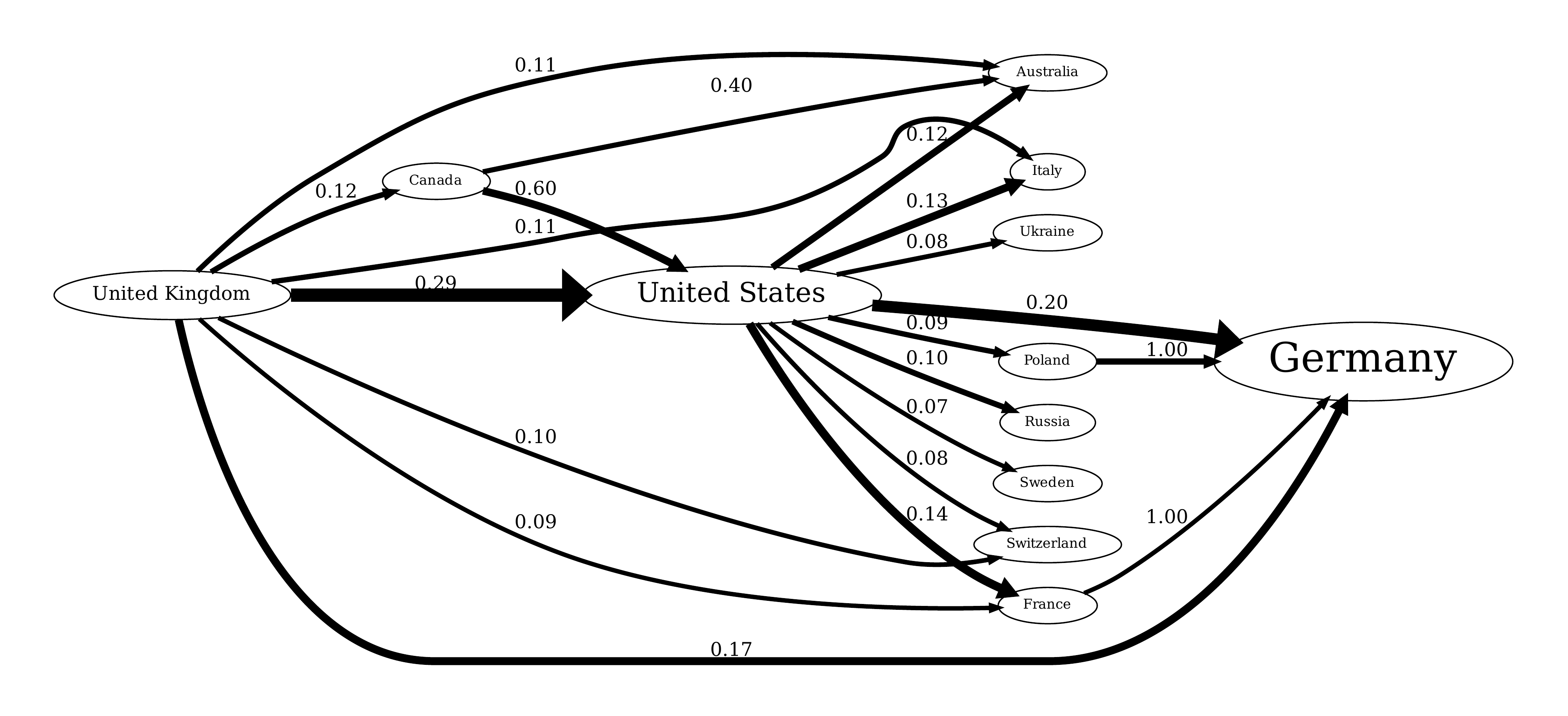}\label{fig:clust0_attacks_graph}}
  \hfill
  \subfloat[Probing pattern on Cluster 2.]{\includegraphics[width=0.47\textwidth]{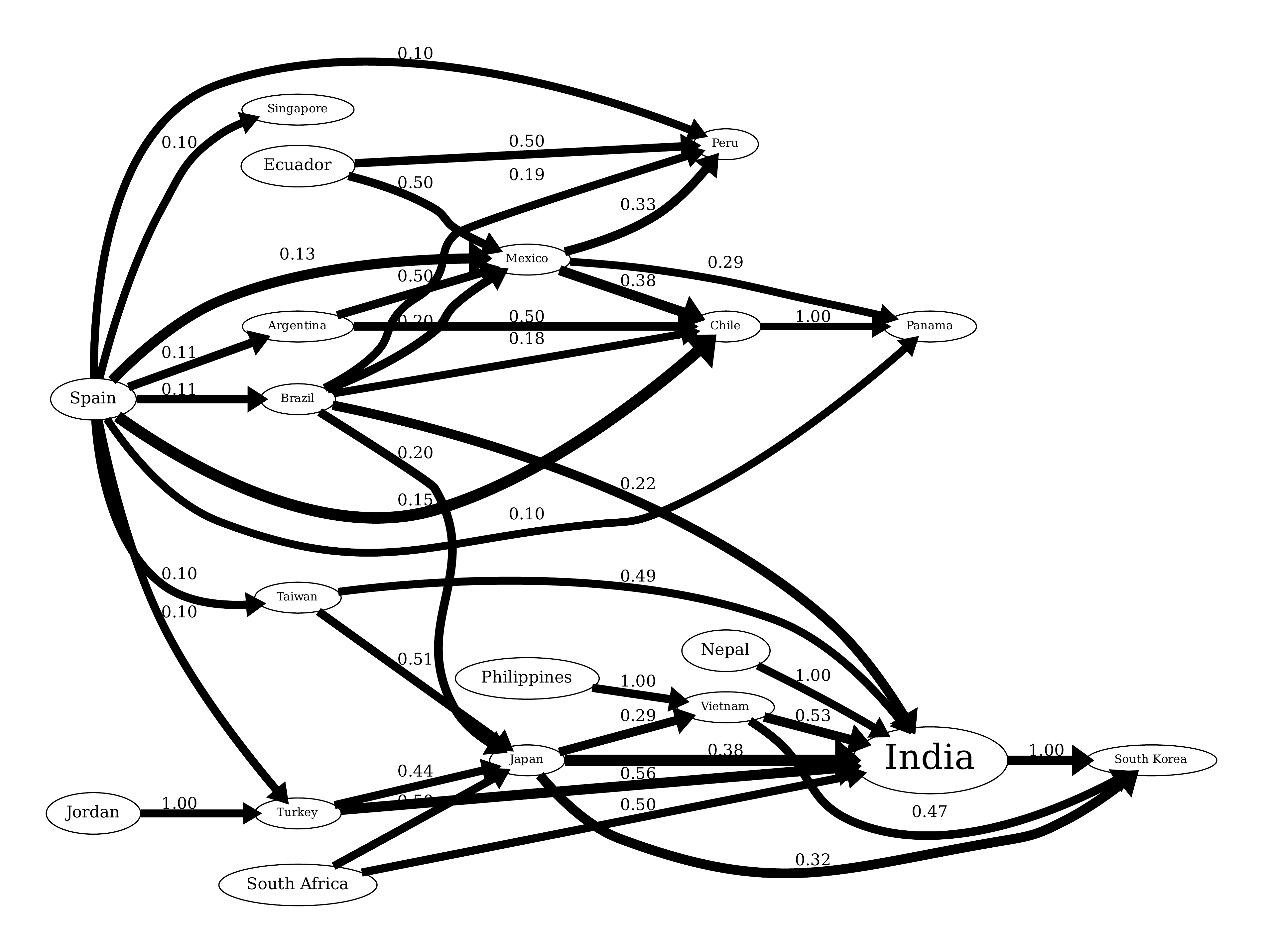}\label{fig:clust2_attacks_graph}}
    \hfill
  \caption{A sample of attacks pattern graphs.
The size of the vertices corresponds to the number of common attacks experienced by targeted countries.}

\label{fig:graphs}
\end{figure}

An attack can be characterized by 3 major elements: targeted country, adversary, and attack type. Our objective is to analyze for each type of attack the geographical propagation behavior. Therefore, we only take into account two characteristics: the type of attack and the country being attacked.
Graphs are generated by aggregating data by type of incident. We then calculate the number of occurrences of countries in the same attack. The totals obtained are then normalized to obtain the probability of occurrence of each tuple given the following transition probability matrix $P$ where $p_{c_ic_j}$ is the probability of an attack to spread from a country $i$ to a country $j$ :
\begin{equation}
    \label{transition_probability_matrix}
    P=
    \begin{bmatrix}
p_{c_1c_1} &  p_{c_1c_2}& ...  & p_{c_1c_n}\\ 
p_{c_2c_1} &  p_{c_2c_2}& ...  & p_{c_2c_n}\\ 
 .&  .&  .&   .\\ 
 .&  .&  .&   .\\ 
p_{c_nc_1} &p_{c_nc_2}  &  ... &  p_{c_nc_n} 
\end{bmatrix}
\end{equation}
Thus $p_{c_ic_j}\geqslant 1 $ and for all $i$, we have
\begin{equation}
    \label{}
    \sum_{c_{k=1}}^{n}p_{c_ic_k} = \sum_{c_{k=1}}^{n}P(c_{m+1}=c_{k}|c_{m} = c_{i})
=1
\end{equation}
This is because, given that an attack targeted a country $c_{i}$, the next country targeted must be one of the possible states. Thus, when we sum over all the possible values of $k$, we should get one. Hence, the sum of each row of the transition probability matrix must be $1$.\\
For each incident i our transition graph G\textsubscript{i}(V\textsubscript{i}, E\textsubscript{i}), in which V\textsubscript{i} is the set of countries targeted by incident i and E\textsubscript{i} represents the probabilities of these countries appearing during a same incident. 
Figure \ref{fig:graphs} shows a sample of attacks occurrence's graph. Figure \ref{fig:clust0_attacks_graph} and Figure \ref{fig:clust2_attacks_graph} represent the probability of an attack to target two countries from the same country.  The two graphs differ from each other by two main components: the number of node that corresponds
to the targeted country and the number of edges describing the
attack behavior. The two graphs bring to light two different attack's pattern. We see that for cluster 0 the number of targeted country  is lower than cluster 1 and the probability of two countries to be attacked together is higher. This can mean that countries of cluster 0 are victim of targeted attacks, aiming to hurt the same countries each time. In the other hand cluster 2 countries may be more concerned by wilder campaigns such as spamming campaigns or probing (i.e, broad attacks campaigns). 

\section{Self tuned country clustering and optimal number of clusters}\label{Clusterin}
In this section, we do the clustering of the countries that have been identified as having been subjected to a cyber security incident. We do so to use this clustering to identify homogeneous structures of countries sharing the same properties. 

We choose to use Spectral clustering over K-means because of the nature of our data. The goal of spectral clustering is to cluster data that is connected but not necessarily compact or clustered within convex boundaries. While k-means is more appropriate for data close to each other (from an Euclidean point of view).

\subsection{Spectral Clustering Result}

\begin{figure}[ht]
\includegraphics[width=\linewidth]{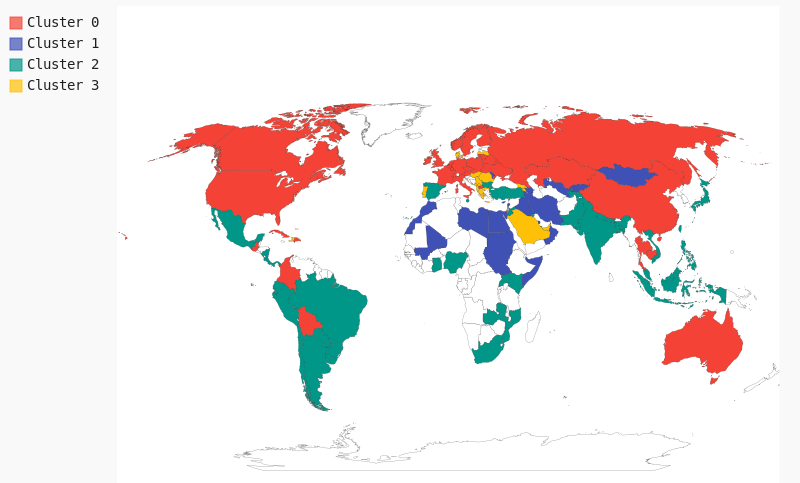}
\caption{Countries Clustering using Spectral Clustering}
\label{fig:map_cluster}
\end{figure}

It is clear from this clustering that the countries belonging to the same cluster are very close geographically. This follows the logic that mass attacks generally target groups of border countries. There is geographical logic behind the clustering that appear from the results shown in Figure  \ref{fig:map_cluster} for example : France, UK, Italy , Sweden and Norway or Morocco, Tunisia, Egypt and Sudan. We also see that Russia appeared alone in its cluster and the USA and Canada appeared in the same cluster as the great powers of the EU. Also the countries of Eastern Europe are grouped together in a same cluster. The same is true for Latin American countries. These results add more evidences to our hypothesis that it is possible to highlight a geographical pattern of cyber attacks. 

\subsection{Threats Time Series correlation analysis} 

\begin{figure}[!ht]
\hfill
\subfloat[Pearson r absolute value heat map of countries time series attack]{\includegraphics[scale=0.3]{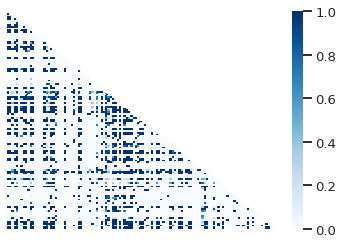}\label{fig:corr_matrix}}
\hfill
\subfloat[Pearson r absolute value heat map using a cross-correlation of 7 days widow size]{\includegraphics[scale=0.3]{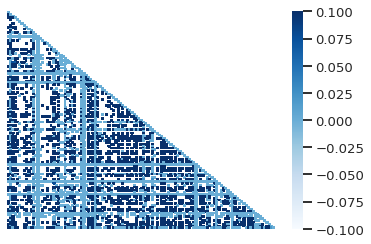}\label{fig:cross_pearson}}
\hfill
\caption{Comparison of point-wise correlation and cross-correlation}
\end{figure}

The last step of the exploratory analysis is to investigate whether there is a pattern concerning the time factor in the propagation of attacks. For this reason we have looked to the correlation between the time series of attacks of different countries by computing the Pearson Correlation Coefficient. The more the two time series are correlated, the closer the Pearson value will be to 1 or -1, and the less they will be, the closer the value will be to 0. Figure \ref{fig:corr_matrix} shows the heat map of the absolute value of Person correlation for all countries listed. We can see that most of them are not very correlated. And those correlated belong to the same cluster. On the other hand, Figure \ref{fig:cross_pearson} shows that by using a 7 days time lag, correlation between the attacks is on clearly increasing for most of the countries. This indicates that in addition to the geographical link ,there is also a considerable link between countries concerning the occurrence of security incidents with a time lag of 7 days.

\section{Covid-19 Case Study}\label{covid-19}

Since the beginning of the covid-19 pandemic in late 2019 and with the resulted curfew and lockdown policies, many activities moved to operate online. This led to a dramatic increase in cyber attacks targeting both individuals and organizations \cite{lallie2021cyber}. In this section we study the threats shared between January 2020 and June 2021 and look at different aspects that characterized the covid-19 era. 

\subsection{Shared threats during Covid-19}

\begin{figure}[ht]
    \centering
    \includegraphics[width=\linewidth]{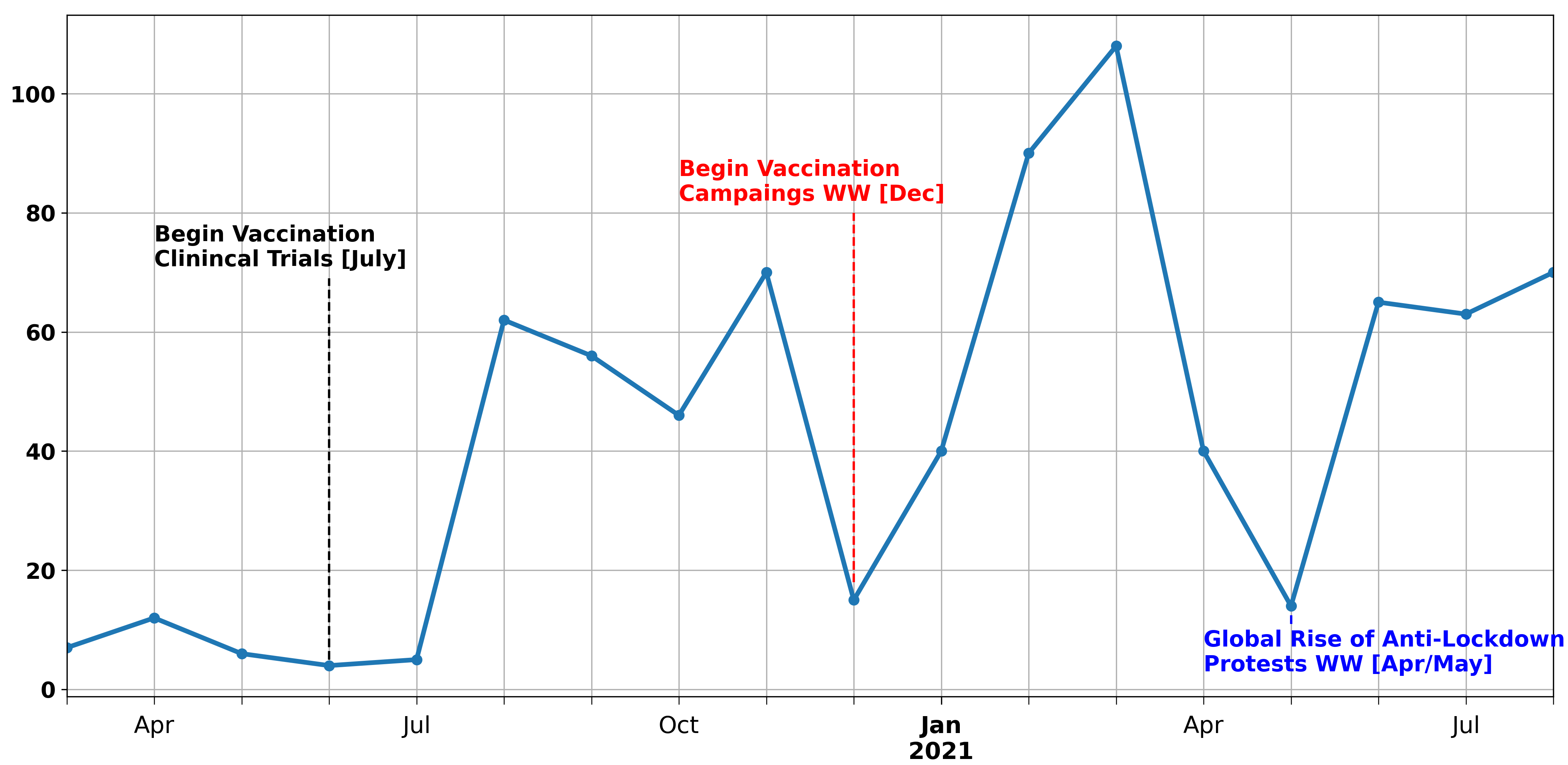}
    \caption{Covid related events and OTX shared threats between Jan 2020 and Aug 2021}
    \label{covid_ts}
\end{figure}

With the pandemic, governments imposed many sanitary measures to limit the spread of the virus or begin the production of the vaccine. As Fig. \ref{covid_ts} illustrates, we observe an increase in cyber threats after the launch of the vaccine clinical trials, after the vaccination campaign have begun and recently the increase could be attributed to the anti-lockdown protests. This shows that covid-19 and its different events and measures had an impact on the increase of cyber attacks around the world. 

\subsection{Attacks targeting vaccine development}

Since the launch of the research about the vaccine development in many countries, particularly in the USA, Germany, UK, etc. threats witnessed an increase. This increase was mainly attributed to APT29 \footnote{https://www.ncsc.gov.uk/news/advisory-apt29-targets-covid-19-vaccine-development} group targeting mostly countries invloved in vaccine developement in order to spy on the production process and breach development data. In this section, we focus on APT29,  analyze its related attack vectors and characteristics. 
MITRE ATT\&CK attributed several attacks ids as techinques used by APT29 \footnote{https://attack.mitre.org/groups/G0016/}. We used these techniques ids to filter only shared threats related to APT29. Doing so, we look for top malware families used by APT29, affected industries and targeted countries. Table \ref{apt29_top5} shows the top 5 malware families which are mostly penetration and spying malwares. In addition, government and manufacturing featured as the most targeted sectors as they are related directly to the vaccine development. As for the top targeted countries, we found that these countries are among the first to begin vaccine research and production.

\begin{table}[ht]
     \centering
     \caption{Top 5 indicators featured in APT29 and related attacks}
     \label{apt29_top5}
     \begin{tabular}{|c|l|}
     \hline 
       Category & Top 5  \\ \hline \hline
       \multirow{5}{*}{\makecell{Malware}} &  Cobalt Strike (Penetration tool)  \\
       & TrickBot (Stealing trojan) \\ 
       & XMRig (cryptominer - trojan-ransom) \\
       & Ryuk (ransomware) \\
       & Agent Tesla - S0331 (spyware) \\
       \hline
       \multirow{5}{*}{\makecell{Industry}} &  Government \\
       & Finance  \\
       & Manufacturing \\
       & Defense \\ 
       & Technology \\
       \hline
       \multirow{5}{*}{\makecell{Country ranking}} &  United States of America \\
       & Germany \\
       & Republic of Korea  \\
       & India \\ 
       & China and United Kingdom \\
       \hline
     \end{tabular}
 \end{table}

\begin{figure}[ht]
\includegraphics[width=\linewidth]{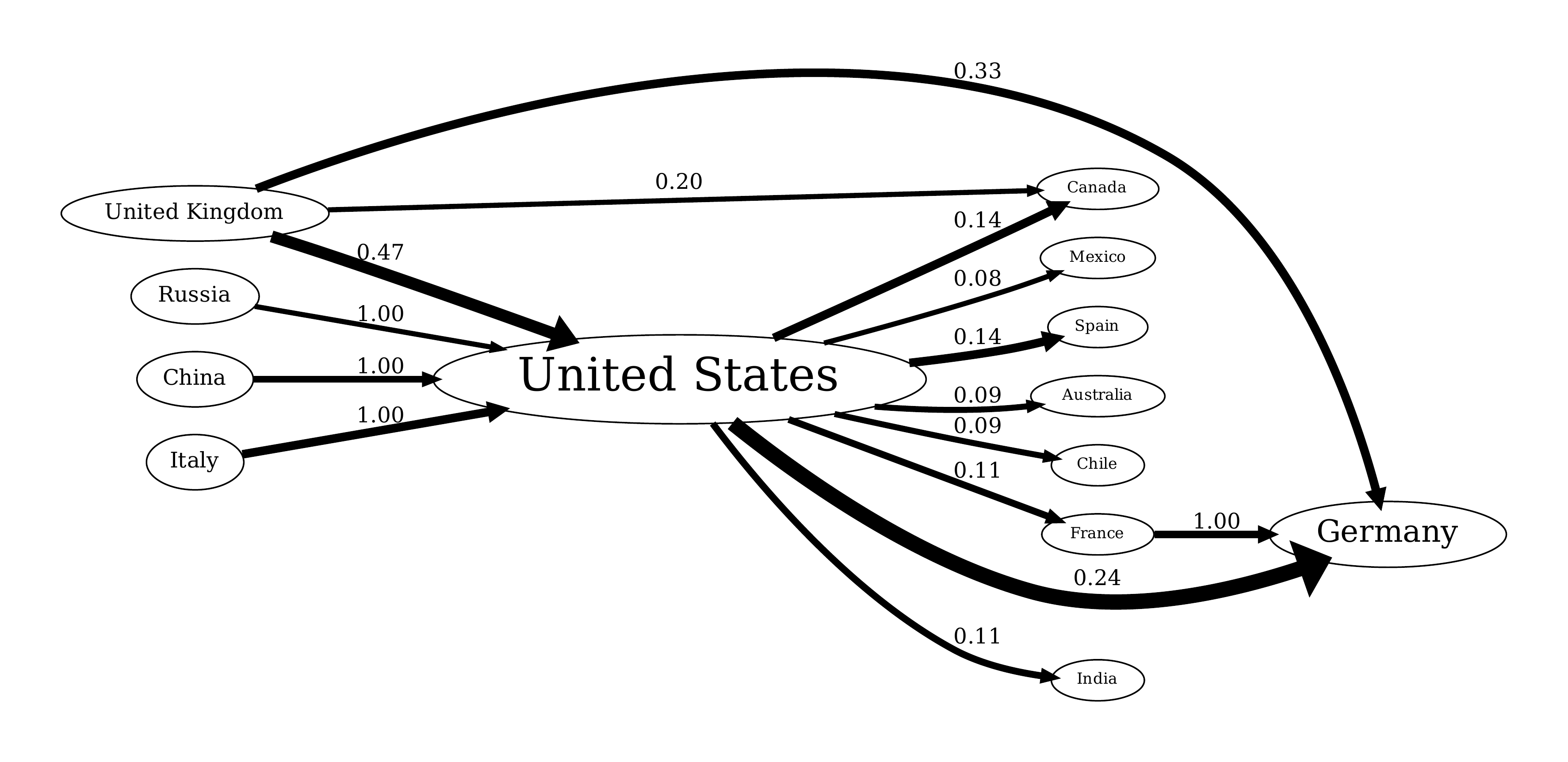}
\caption{Propagation of APT29 related attacks during COVID-19 Pandemic}
\label{fig:APT29_graph}
\end{figure}

Moreover these countries are being targeted usually together by the same attacks, as shown by the high probabilities of transition in Figure \ref{fig:APT29_graph}. We can see for example that 47\% of the cases where UK (Astrazeneca) have been targeted by APT29;, USA (Pfizer, Moderna and Jansen) have been targeted too. Also in 33\% of the cases Germany(Pfizer) is targeted when UK is. Also the fact that few countries have been targeted shows that this attacks are targeted attacks and not broad campaign.

\section{Conclusion}
\label{conlusion}

This work presented an exploratory data analysis performed on 155k shared threats including 10k cyber threat related to COVID-19. The investigation of this data has highlighted the existence of a spatio-temporal trend of cyber threats between countries and that they can be classified into clusters. It also showed that there is an evolution over years on top malware types used. Our study showed that abnormal event relative to COVID-19 appear to have a loose correlation with the increase of cyber-attack campaigns. Our research presents opportunity for further investigations.
This research has shown what can best be described as a loose direct relationship between COVID-19 related events and cyber-attacks. Further research should investigate this phenomenon and outline whether a predictive model can be used to confirm this relationships and patterns.

\bibliographystyle{IEEEtran}
\bibliography{Bibliography}

\end{document}